\documentclass[12pt,preprint]{aastex}
\newcommand{\kms}{\mbox{km\,s$^{-1}$}}

\shorttitle{Chemically tagging the HR1614 moving group}
\shortauthors{De Silva et al.}

\begin{document}

\title{Chemically tagging the HR1614 moving group}


\author{G.M. De Silva\altaffilmark{1} and K.C. Freeman}
\affil{Mount Stromlo Observatory, Australian National University,
Weston ACT 2611, Australia}
\email{gayandhi@mso.anu.edu.au}

\author{J. Bland-Hawthorn}
\affil{Anglo-Australian Observatory, Eastwood NSW 2122, Australia}
\email{jbh@aao.gov.au}

\author{M. Asplund and M.S. Bessell}
\affil{Mount Stromlo Observatory, Australian National University,
Weston ACT 2611, Australia}

\altaffiltext{1}{Now at European Southern Observatory, Alonso de Cordova 3107, Casilla 19001, Santiago 19, Chile}



\begin{abstract}
We present abundances for a sample of F,G,K dwarfs of the HR1614 moving group based on high resolution, high S/N ratio spectra from AAT/UCLES. Our sample includes stars from \citet{FH}, as well as from \citet{e98}. Abundances were derived for Na, Mg, Al, Si, Ca, Mn, Fe, Ni, Zr, Ba, Ce, Nd, and Eu.  The alpha, Fe, and Fe-peak element abundances show a bimodal distribution, with four stars having solar metallicities while the remaining 14 stars are metal rich [Fe/H] $\ge$ 0.25
dex. However the abundances of these two groups converge for the heavier n-capture elements. Based on their photometry and kinematics, three of the four deviating stars are likely non-members or binaries. Although one star cannot be excluded on these grounds, we do expect low-level contamination from field stars within the HR1614 moving group's range of magnitude, color and space velocities. \\

 Disregarding these four stars, the abundance scatter across the group members for all elements is low. We find that there is an 80\% probability that the intrinsic scatter does not exceed the following values: Fe 0.01; Na 0.08; Mg 0.02; Al 0.06; Si 0.02; Ca 0.02; Mn 0.01; Ni 0.01; Zr 0.03; Ba 0.03; Ce 0.04; Nd 0.01 and Eu 0.02 dex. The homogeneity of the HR 1614 group in age and abundance suggests that it is the remnant of a dispersed star-forming event.  Its kinematical 
coherence shows that such a dispersing system need not be significantly 
perturbed by external dynamical influences like galactic spiral structure
or giant molecular clouds, at least over a period of 2 Gyr.


\end{abstract}


\keywords{Galaxy: evolution --- Galaxy: open clusters and associations: individual(HR1614 moving group) --- stars: abundances}



\section{Introduction}
The concept of moving groups was first introduced by Olin Eggen in the 1960s. Basically the stars form from a common progenitor gas cloud. As the cluster disperses around the Galaxy, it stretches into a tube-like
structure around the Galactic plane and after several Galactic orbits it
will dissolve into the Galactic background. If the Sun
happens to be inside this tube, the  member stars will appear all over the sky but may be identified as a group through their common space velocities.  A moving group is therefore the in-between step from bound clusters to field stars.\\

For this reason moving groups are an important class of objects for the purpose of testing the chemical tagging technique put forward by \citet{fbh02}. In summary, the long term goal of chemical tagging is to reassemble the ancient star-forming aggregates in the Galactic disk by studying their chemical signatures, in order to unravel the sequence of events involved in the dissipative formation of the disk. One of the fundamental requirements for the viability of chemical tagging is chemical homogeneity within individual Disk clusters. Since most bound open clusters are believed to be chemically homogeneous (eg. Hyades \citep{desilva} and Collinder 261 \citep{cr261}), establishing homogeneity in unbound moving groups which retain some kinematical identity is the next step for the chemical tagging technique. The final step is to chemically identify groups which have no dynamical identity. \\

Although the existence of many stellar moving groups has been
suggested, the reality of most is yet to be verified. 
In the past due to the lack of accurate parallaxes it has been
difficult to reliably identify the stellar members of moving groups. 
This has now been partly overcome from the observations by the
Hipparcos satellite. The presence of a moving group associated with the star
HR1614, first advocated by \citet{e78a},
was verified by \citet{FH} using the new
Hipparcos parallaxes and recent radial velocities, and
provides an extended sample of stellar members. \\

Eggen's initial identification of the moving group came from studying a
sample of stars within $\pm$ 10\kms\ of the galactic rotational velocity (V) of star HR1614. He used the high excess in $b-y$ as a membership criterion for his
stars to obtain a color-magnitude diagram \citep[Fig 1a.]{eggen78} resembling
that of an old open cluster. Based on a few spectroscopic studies and
positions in the M$_{V}$ vs.~$R - I$ color-magnitude diagram, \citet{eggen78}
concluded that the majority of stars he classified as members were as metal
rich as the Hyades.\\

Followup studies by \citet{smith83} using DDO photometry showed enhanced
cyanogen bands in \citet{e78a} sample stars. The
derived metallicity confirmed Eggen's estimate of a high metallicity, but
two giants found to have significantly low metallicity cast doubt
over Eggen's $b-y$ membership criteria. Enforcing a strict criterion
using UBV, DDO and $b-y$, \citet{smith83} showed that many of Eggen's original
candidates did not belong to the group. \citet{e92}, utilizing the
CN-enhancement, provided a second sample of main sequence dwarfs with the same
V-velocity restrictions as applied earlier. With the accurate parallax
measurements made available from the Hipparcos mission, \citet{e98} 
showed that the group membership for his HR1614 sample \citep{e92} is supported by Hipparcos parallaxes. However for two stars where the discrepancies were large, \citet{e98} suggested errors in the Hipparcos parallaxes.  \\

\citet[hereafter FH]{FH} utilized Hipparcos parallaxes and radial velocities from several
recent catalogues to perform an unbiased search for
HR1614 member stars over a large region of UV space. They conclude
that there is a distinct stellar population of metal rich stars
centered at U = 10 \kms, V = $-60$ \kms\ and tilted in the UV plane, which they  
associate with the HR1614 moving group. Following simulations by \citet{skul}
and their own basic simulations, FH find the observed tilt in the UV plane
to be a feature of a moving group. In contrast, the classical criteria used by
Eggen (based on epicyclic theory), requires all member stars to lie at constant V in the UV plane. FH state that such central clumping will only occur if the Sun is
located very close to the center of the moving group, otherwise a tilt is to
be expected.\\ 

With the past literature pointing towards its reality, the HR1614
moving group is an attractive target for testing chemical tagging. The star cluster is quite distinctive due to its age of about 2 Gyr, and high metallicity of [Fe/H]= +0.25. Its the stellar motions suggest that these stars formed in the inner disk.
\citet{dehnen99} \citep[see also][]{raboud}, shows that there is an overdensity in the UV plane at U = -20 \kms\ and V = -45 \kms\ (earlier named the U-anomaly), which he associates with stars thrown out from the inner disk, an effect particularly strong close to the Galactic bars' outer Lindblad resonance, where the Sun is located outside the outer Lindblad resonance. Its likely origin in the inner disk makes the HR1614 moving group of particular interest. It gives us a unique chance to test chemical tagging precepts on stars that originated away from the solar circle. The HR1614 moving group has never been subject to a high resolution abundance study. We have now for the first time performed a detailed abundance analysis.\\

\section{Observations}
The original data were obtained at the 4m AAT, using UCLES in August 2003.
Observations were done at two wavelength settings at the blue and red. The blue
setting was centered at 4130\AA\ and covers the region from 3800\AA\ to 4700\AA, while the red
setting was centered at 7040\AA\ covering the spectral range from 5510\AA\ to
10200\AA. The 31.6 lines mm$^{-1}$ echelle grating was always used. The slit
was opened to a width of 1.2 arcsecs on the sky providing a resolving power of
48,000. \\

The observing routine included 10 bias
frames, and 5 quartz lamp exposures to provide data for flat fielding, taken at
the start of each night. A Th-Ar hollow cathode lamp spectra was taken
before and after each stellar exposure. Two radial velocity standards, one at
the start and one at the end of each night, were also observed. The stellar
exposures were taken in batches of at least 3 exposures per star, with each
exposure typically around 200 secs. All program stars have a total signal-to-noise ratio greater than 150 per pixel at the central wavelengths. \\

The spectroscopic data were reduced using the IRAF packages IMRED, CCDRED, and
ECHELLE. The preliminaries included biasing, flat fielding, scattered light
removal, order extraction, wavelength calibration and continuum
fitting. Details of the final sample selected for abundance analysis are given in Table
\ref{tab:sample}.\\

\section{Abundance analysis}\label{c5sec:ab}

\subsection{Model Atmospheres and Spectral Lines}
The abundance analysis makes use of the latest version of the MOOG code
\citep{sneden73} for LTE EW analysis and spectral syntheses.
Interpolated Kurucz model atmospheres based on the ATLAS9 code
\citep{cas97} with no convective overshoot were used throughout this study.\\

Now we focus on the lines of Na, Mg, Al, Si, Ca, Mn, Ni, Fe, Zr,
Ba, Ce, Nd, and Eu. A full line list is given in Table \ref{hr1614:lines} as well as EWs of the reference star. The $gf$
values for the detected lines of Na, Mg, Al, Si, Ca, Ni, and Zr were obtained from
a combination of lines from \citet{ap04,yong05,reddy03} and \citet{P03}. For
Mn, the $gf$ values were taken from \citet{prochaska00} and include the effects
of hyperfine splitting. The main sources of the Fe~{\sc i} line data is the
laboratory measurements by the Oxford group 
(Blackwell et al., 1979a,b, 1995 and references therein). This was supplemented
by additional lines from \citet{reddy03}. For Fe {\sc ii} we adopt the $gf$ values from \citet{biemont91,P03} and \citet{fe2lines}. The only measurable Zr line
data was obtained via the Viena Atomic Line Database (VALD\footnote{http://ams.astro.univie.ac.at/vald/}; \citet{VALD,VALD1,VALD2,VALD3}), from the Bell heavy database (Kurucz CD-ROM 18, 1993). Ba $gf$  values were adopted
from \citet{prochaskaba} as well as \citet{ap04}. The Ce $gf$ values
were obtained from the Database on Rare Earths At Mons University (DREAM\footnote{http://w3.umh.ac.be/~astro/dream.shtml}; \citet{dream}). The $gf$ values for Nd were taken
from \citet{dlsc03}, and finally, the Eu $gf$ values were taken from \citet{lawler01} and we account for hyperfine splitting and isotopic shifts assuming a
solar isotopic mix. \\

\subsection{Stellar parameters}
\label{c5:stellarparams}
\begin{figure}
\begin{center}
\includegraphics[scale=0.7, angle=0]{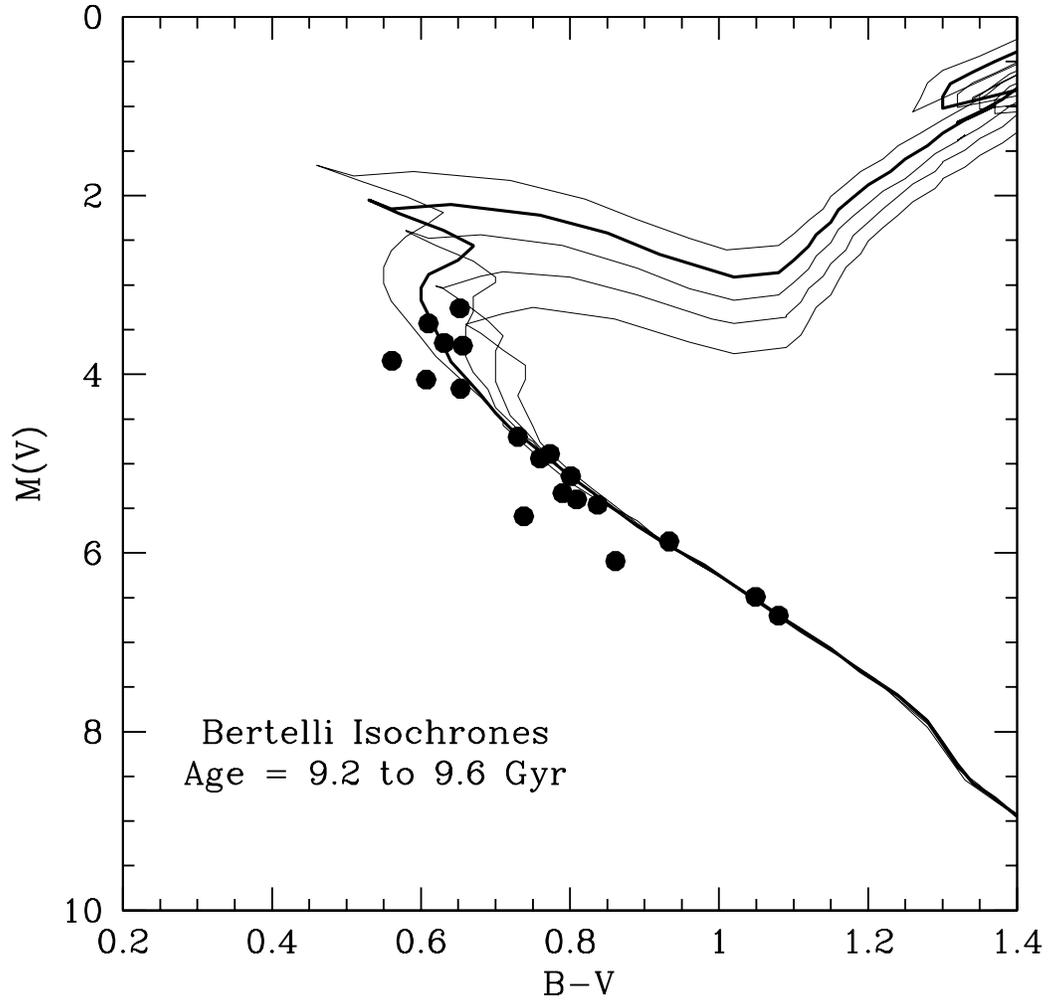}
\caption{Our sample stars overlaid with \citet{bert} isochrones. The best
  fitting isochrone of $\approx$ 2 Gyr is highlighted. Note all
  isochrones are for Z = 0.05.}\label{fig:iso}
\end{center}
\end{figure}

Initially T$_{eff}$ was estimated using the color-temperature relations of
\citet{alonso96}, equation (7) and (8), using V-K values where the K magnitudes were obtained from 2MASS and adopting [Fe/H] = +0.2. These temperature
estimates were used as an initial guess when deriving spectroscopic temperatures. An estimate of log $g$ was then obtained using the distances (from Hipparcos parallaxes), bolometric corrections, and stellar mass from \citet{bert}
isochrones. The best fitting isochrone was found to be of age $\approx$ 2 Gyr,
consistent with the estimate obtained by FH. Figure \ref{fig:iso} shows our sample of stars overlaid with several isochrones for Z = 0.05.\\ 

Next we derived the stellar parameters based on spectroscopy. We compute
abundances for all Fe {\sc i} and Fe {\sc ii} lines based on the measured EW. The EWs were
measured by fitting a Gaussian profile to each line  using
the interactive $SPLOT$ function in IRAF. T$_{eff}$ was
derived by forcing the Fe {\sc i} abundances to be independent of excitation potential, ie. excitational equilibrium. Microturbulence was derived from the condition
that Fe {\sc i} lines show no abundance trend with EW. Log $g$ was derived via ionization
equilibrium, ie. the abundances from Fe {\sc i} equals Fe {\sc ii}. The photometric
estimates of T$_{eff}$ and log $g$ were used as the initial guess model and iterated
until a self consistent set of model parameters were obtained. The spectroscopic  estimates of T$_{eff}$ are hotter than the photometric estimates on average by 150 K. For our abundance analysis we adopt the spectroscopic stellar parameters.  Table \ref{tab:params} lists the adopted parameters. \\

\subsection{Elemental Abundances}\label{sect:abundances}
Depending on the degree of blending of the spectral lines, abundances were 
derived either by EW measurements or by spectral synthesis. The EW
measurements were used for abundance determinations for all elements, except
the heavy elements from Ba to Eu, which required full spectral synthesis. \\

\begin{figure}
\begin{center}
\includegraphics[scale=0.7, angle=0]{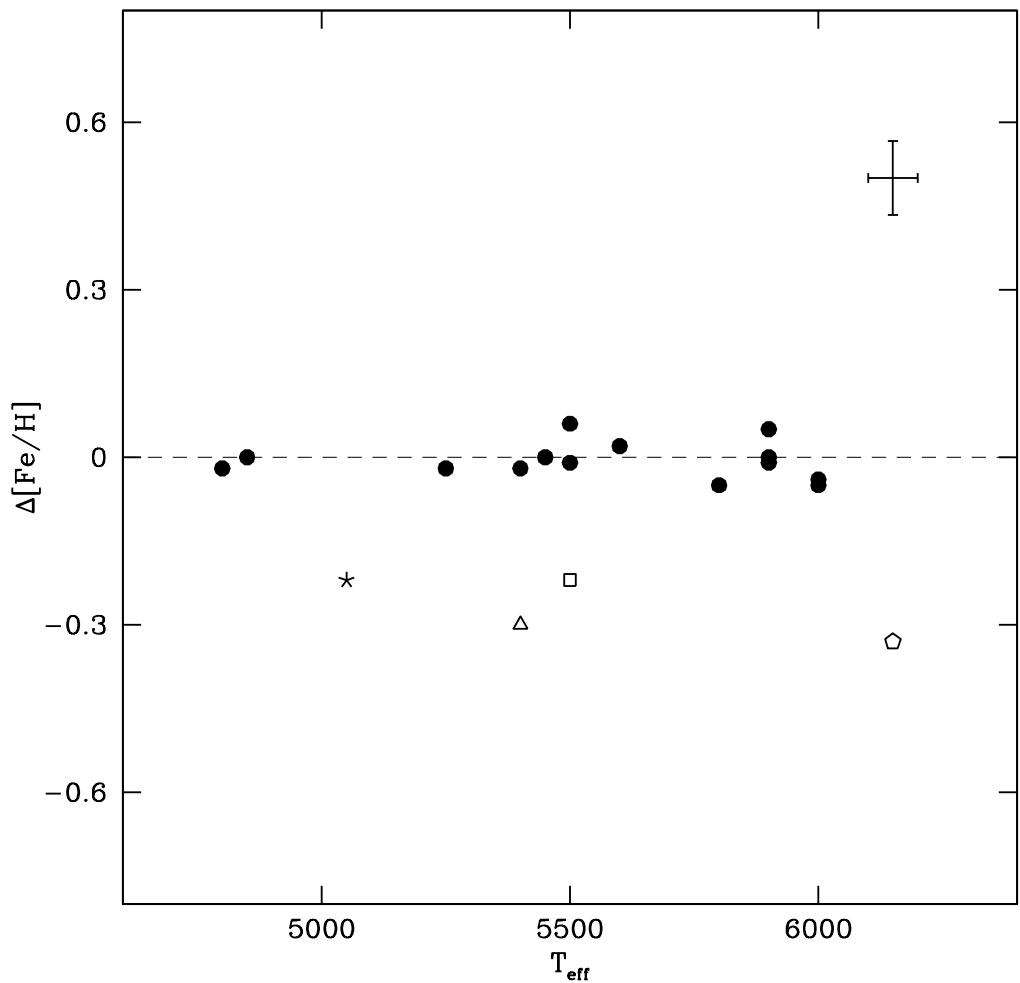}
\caption{Differential Fe abundances. Typical errors are shown in the top
  right. Note that the y-axis zero level corresponds to
  the abundance of star HR1614, and does not represent the solar level as our
  abundances are calculated relative to HR1614. The majority of the sample
  (filled circles) are metal rich with [Fe/H] $\approx$ 0.25 dex. The lower
  four stars have solar metallicities. These stars are HIP13513 (star),
  HIP6762 (triangle), HIP25840 (hexagon) and HIP116970 (square). The
  membership probabilities of these stars are discussed in Section
  \ref{sect:membership}. }\label{fig:fe}
\end{center}
\end{figure}
\begin{figure}
\begin{center}
\includegraphics[scale=0.7, angle=0]{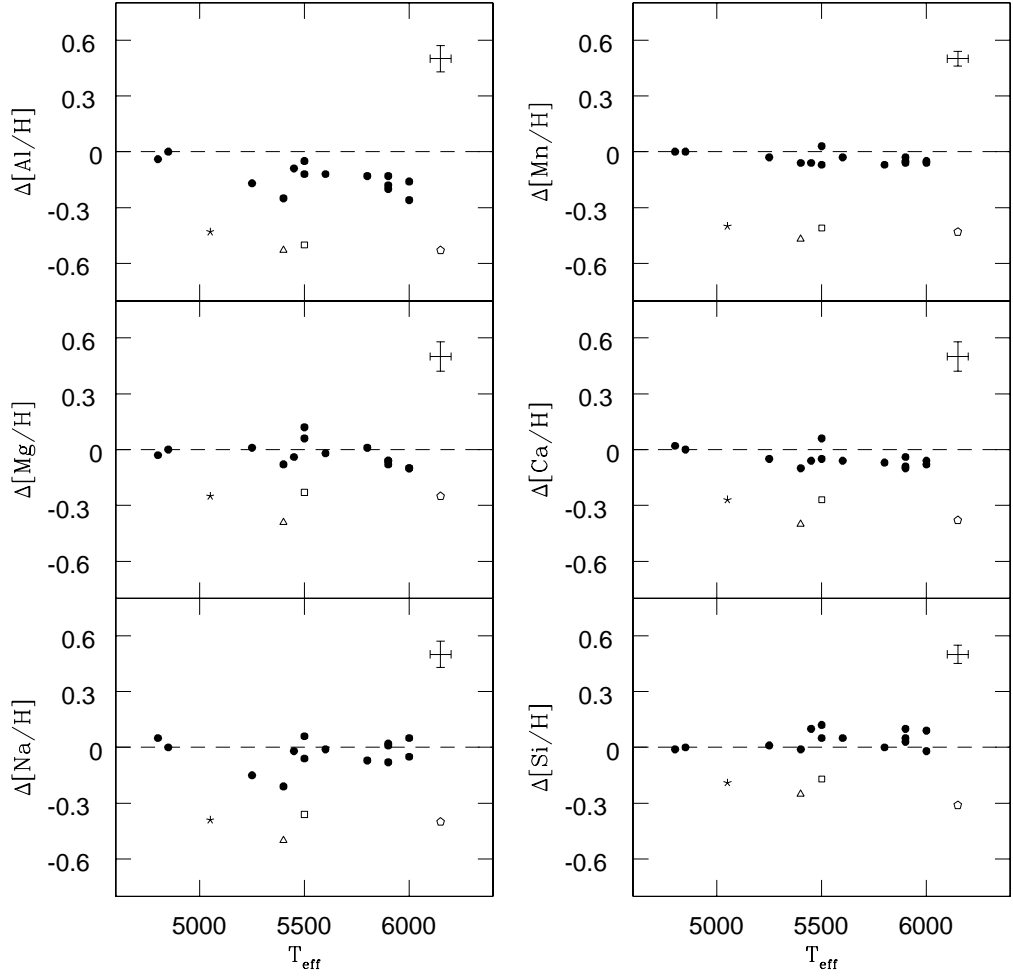}
\caption{Differential abundances relative to HR1614 for elements up to Mn. The symbols are the same as those for Fig \ref{fig:fe}. The
  bimodality seen in Fig \ref{fig:fe} is clearly present in these elements
  as well.}\label{fig:alpha}
\end{center}
\end{figure}
\begin{figure}
\begin{center}
\includegraphics[scale=0.7, angle=0]{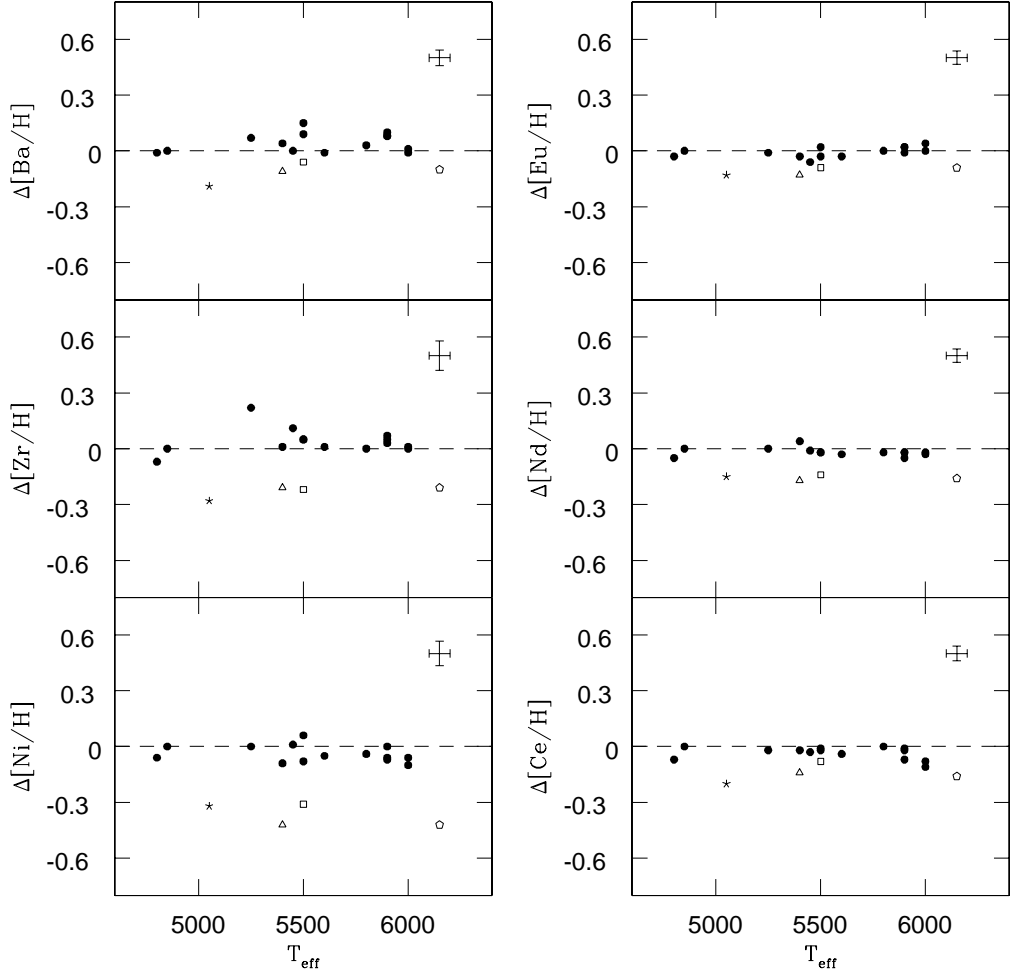}
\caption{Differential abundances relative to HR1614 for elements heavier than
  Fe. The symbols are the same as those for Fig \ref{fig:fe}. While a clear
  bimodality is observed for Ni and Zr, the two groups tend to
  converge for the heavier s- and r- process elements.}\label{fig:heavy} 
\end{center}
\end{figure}

In our initial analysis, we derived absolute abundances, but for the purposes of testing chemical tagging, we decided to use differential abundances as these were found to have smaller systematic errors. The final differential abundances $\Delta$[X/H] were derived by
subtracting the absolute abundance of each individual line of the reference
star HR1614 from the same line in the sample stars, and then taking the mean of
the differences for each element. By using such a line-by-line differential
technique we can identify and remove any problematic lines, reducing the errors due to the uncertainty in the line data,
hence minimizing the star-to-star scatter. The differential Fe abundances
$\Delta$[Fe/H] is plotted in Fig \ref{fig:fe}. The differential
abundances for all elements lighter than Fe are plotted in Fig
\ref{fig:alpha} and those heavier than Fe are plotted in Fig
\ref{fig:heavy}. Note that the zero level in these plots do not
correspond to the solar metallicity, as these are plots of the differential
abundances relative to HR1614. Also, note that four stars (HIP 13513, HIP6762,
HIP 25840, HIP 116970) were found to have solar-level abundances for all
elements except for the heavier n-capture elements. These stars are
identified in the plots with different symbols. We will discuss our results
and these deviating stars in detail in Section \ref{sect:discuss}. The derived
absolute abundances (log $\epsilon$) of the sample stars are presented in Table
\ref{tab:abundance}. 

\subsection{Error Analysis}\label{hr1614errors}
The main sources of error in the present study are that of EW
measurements, continuum placement and stellar parameters. Because we are only interested in the differential abundances, external
errors, such as uncertainties in the line data and model atmospheres are the
least sources of error. The number of lines used to calculate the final abundances also contributed to the total uncertainty for each element.\\

The error in EWs was estimated by repeated measurements of each line. The measurement errors for the synthesized abundances were derived by
changing the abundance until there is a clear visible deviation from the best
fit. The error in the stellar parameters were estimated to be
 $\delta T_{eff}$ = 50K, $\delta$log \emph{g} = 0.1 cm\,s$^{-2}$ and
$\delta \xi$ = 0.1 \kms\ based on our spectroscopic derivation . Table \ref{tab:sens} summarizes the abundance
sensitivities to each stellar parameter and methods of derivation for two stars at either end of the temperature range. Typical values of the final error is as follows; Fe = 0.05 dex, Na = 0.07 dex, Mg = 0.07 dex, Al = 0.07 dex, Si = 0.05 dex, Ca = 0.07 dex, Mn = 0.04 dex, Ni = 0.06 dex, Zr = 0.07 dex, Ba = 0.05 dex, Ce = 0.03 dex, Nd = 0.03 dex and Eu = 0.03 dex.
 \\

The four stars that show solar-level metallicities in the lighter elements
were re-assessed to check whether the stellar parameters or other systematic
parameters were at fault. However this was not found to be the case, with
large changes in the model parameters ($\delta$T$_{eff} \ge$ 300 K, $\delta$log $g \ge$ 2, and
$\delta \xi \ge$ 1) needed to bring these star's abundances in line with the
other enriched stars.  
In conclusion, we find that systematic error is unlikely to explain the behavior of these four stars. \\

\section{Dynamical Analysis}

\subsection{Membership Criteria}
While open cluster members are easily determined from radial velocities with errors less than a few kilometers per second, moving group membership is based on UVW space motion, but the exact criteria for selecting members remain somewhat uncertain.\\

Eggen's method of isolating moving group members required all member stars to have the same V-velocity, ie. in the direction of the Galactic rotation. 
In the epicycle approximation, as a star orbits around the Galaxy, it travels on an epicycle about the circular guiding center of its Galactic orbit. If member stars, which would have originated from the one location, are now in the solar neighborhood, within the observable horizon, then this limits their guiding
center radii to be very close to that of the Sun. Those member stars with longer and shorter guiding center radii would be either ahead or lag behind the Sun and therefore not be within a small locally defined volume. This means that HR 1614 group member stars located within a small volume should have a similar V-velocity. Any scatter in V would reflect the scatter in the orbital radii. This variation cannot be large for stars within a small volume.\\

Work by FH and \citet{skul} show that such a clump in the UV space is not
necessary for all members of the moving group. Their modeling shows
that a small $tilt$ in the UV space is the dynamical signature of the moving
group. This seems to be in contradiction with the above mentioned concept, and is primarily due to the limitations of the epicyclic approximation, but it does mean that the true kinematical signature of moving groups remain uncertain. The
sample selected for our chemical analysis includes stars selected from Eggen
(via the V-clump criteria) as well as from the FH sample (selected via the
tilted criteria). As establishing accurate membership is a key component of
our study, obtaining reliable velocity information is essential. While some velocity information is available from the literature, we have decided to undertake our own measurements in order to obtain a consistent set of velocities. \\

\subsection{Radial velocities}\label{c5sec:rv}
Radial velocities (RVs) were determined by Fourier transform cross
correlation of template spectra with observed spectra, making use of the
packages RVSAO / XCSAO \citep{rvsao,xcsao} that are run under
IRAF. From the available spectra, RVs were estimated using the blue region from 4200
- 4400 \AA\ and the red region from 8465 - 8680 \AA\ , although the red region 
containing the calcium triplet was given a higher weighting. Template spectra
from \citet{zwitter} were obtained via private communication from M.
Williams. Since the stellar parameters were already established from our
earlier spectroscopic studies, templates matching closest to the sample
parameters were selected for the cross correlation. \\

The procedure undertaken to determine RV was as follows. The
barycentric velocity correction was applied to all observed stars
using the IRAF package BCVCORR. The data files were then re-sampled
according to the template sampling, which had a resolution of 20,000. The
continuum normalization of the data were rechecked to
ensure almost complete flatness. The template files were cropped to
the wavelength ranges of the observed data. Since the observed echelle
spectra has inter-order gaps in the red region, the red templates were split
into two separate regions according to the available observed spectral
windows. The two regions were from 8465 - 8550 \AA\ and 8593 - 8680 \AA\,
which included all three of the Ca~{\sc ii} triplet lines. For all cross
correlations, the apodizing option in XCSAO was turned on to smooth the edges
(up to 5\% of the spectra) to zero by multiplication with a cosine bell
function. Bandpass filtering was set with the parameters low-bin and top-low
for low frequency cut-offs, and parameters top-run and nrun to filter the high
frequencies. The data was rebinned to 2048 bins set by ncol. By experimentation, the low frequency cut-offs were set to [4,20] which allows the broader spectral features through, while filtering out any residual low frequency structure in the continuum. For the high frequencies, we set the cut-offs to [240,580] after several experimentations to avoid suppressing any line features and ensuring all high signal data passes through, while the high noise regions are removed.\\

Resulting radial velocity values were selected based on the goodness-of-fit
parameter (R) as output from XCSAO. The highest R value for all program stars
were over 20 in the blue spectral region and over 30 in the red. While both
red and blue regions gave very similar results, the final radial velocities are
an average of both the red and blue regions, with a weighting of 3:2 in favor
of the red region based on the R values. The exception to this is the star HIP13513, which we measure to have a variable RV with an RV of 3.5 \kms\ in the red and -46.9 \kms\ in the blue, which was taken at a later date. Note that this star is one of the chemically deviating stars mentioned
in Section \ref{sect:abundances}. We will discuss these stars
later in Section \ref{sect:membership}. With the exception of HIP13513, our results agree better with the measurements by \citet[N04]{nord04} where the average difference is about 0.4 \kms\ , than for FH where the average difference is about 4 \kms\ , although they agree within the given error bars of the FH sources. Our RVs for all
sample stars and RV standards, as well as several literature values, are
listed in Table \ref{tab:rv}. Our errors as calculated by XCSAO are less than 1 \kms. \\

\subsection{U V W velocities}

Using the above computed RVs, and proper motions and parallaxes from Hipparcos, we calculated UVW space velocities making use of
the conversion code obtained via private communication from M.~Williams. The resulting velocities are with respect to the LSR adopting the standard
solar motion of U = 10 \kms, V = 5 \kms, and W = 7 \kms\ \citep{db98}. Table \ref{tab:uvw}
lists our derived velocities, as well as those from FH. Overall our velocities are in good agreement with FH with mean differences of about 1.7, 0.6, and 2.1 \kms\ in U, V and W respectively.\\

\begin{figure}
\begin{center}
\includegraphics[scale=0.75, angle=0]{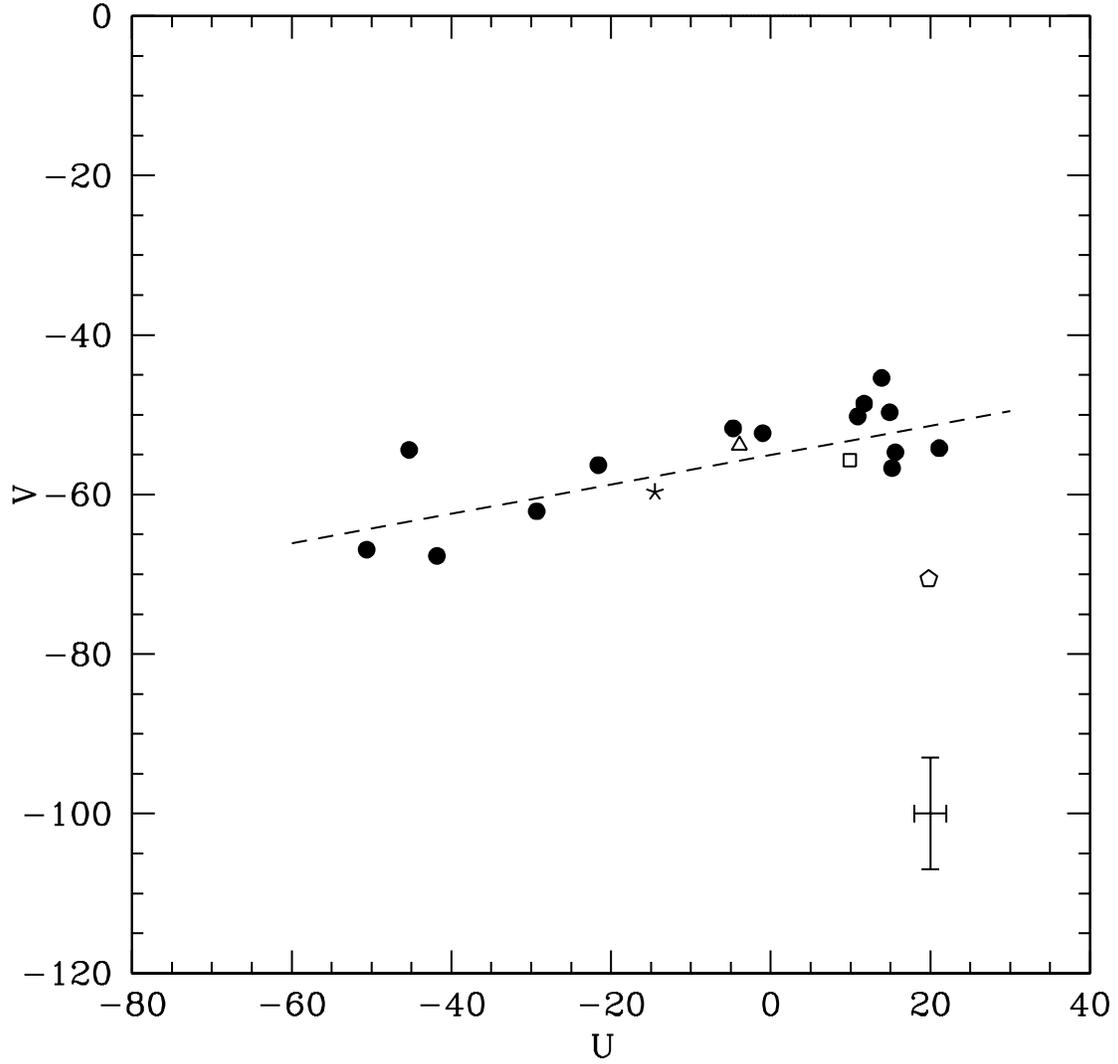}
\caption{The derived UV velocities of the sample stars. Note the chemically
  deviating stars are identified with different symbols as in Fig \ref{fig:fe}. The dashed line represents the line of best fit. Typical errors are shown on the bottom right corner.}\label{fig:uv}
\end{center}
\end{figure}

Figure \ref{fig:uv} shows the position of our stars in the UV plane using the new velocities obtained in this study. The presence of a slight slope seems to support the tilted membership criterion argued by FH. By fitting a least squares regression line of the form, V = a + bU, we find a = -55.06 and b = 0.18. The significance of this slope depends on the measurement errors. Taking into account errors in RV's and Hipparcos
parallaxes, our typical errors are $<$ 2 \kms\ for U, $<$ 10 \kms\ for V and $<$ 5 \kms\ for W. Considering the uncertainty in the V-velocity to be on average 6 \kms\ and disregarding the small uncertainty in U-velocity, we derive the uncertainty in the gradient (b) of the best fitting line to be $\pm$ 0.05.  Therefore the slight tilt in Figure \ref{fig:uv} is marginally significant and lends support to FH's tilted membership criterion in the UV plane. We will discuss the issue of membership in detail in Section \ref{sect:membership}.

\section{Discussion}
\subsection{Chemical Homogeneity}\label{sect:discuss}

We have studied a total of 18 stars that were thought to be members
following the kinematical criteria of FH and \citet{e98}. Out of this sample, four stars
are found to deviate from the cluster mean in all elements except the
n-capture elements. Disregarding these four stars for the moment, our abundance analysis
demonstrates that the moving group is chemically enriched with mean [Fe/H] =
0.25 adopting a solar value of 7.52 \citep{snedenfe}, comparable to the result obtained by FH using photometric
techniques given the estimated errors. Table \ref{tab:scatter} summarizes
the cluster mean abundances for all studied elements and the rms scatter, both
including and excluding the four deviating stars. \\

\begin{figure}
\begin{center}
\includegraphics[scale = 0.65]{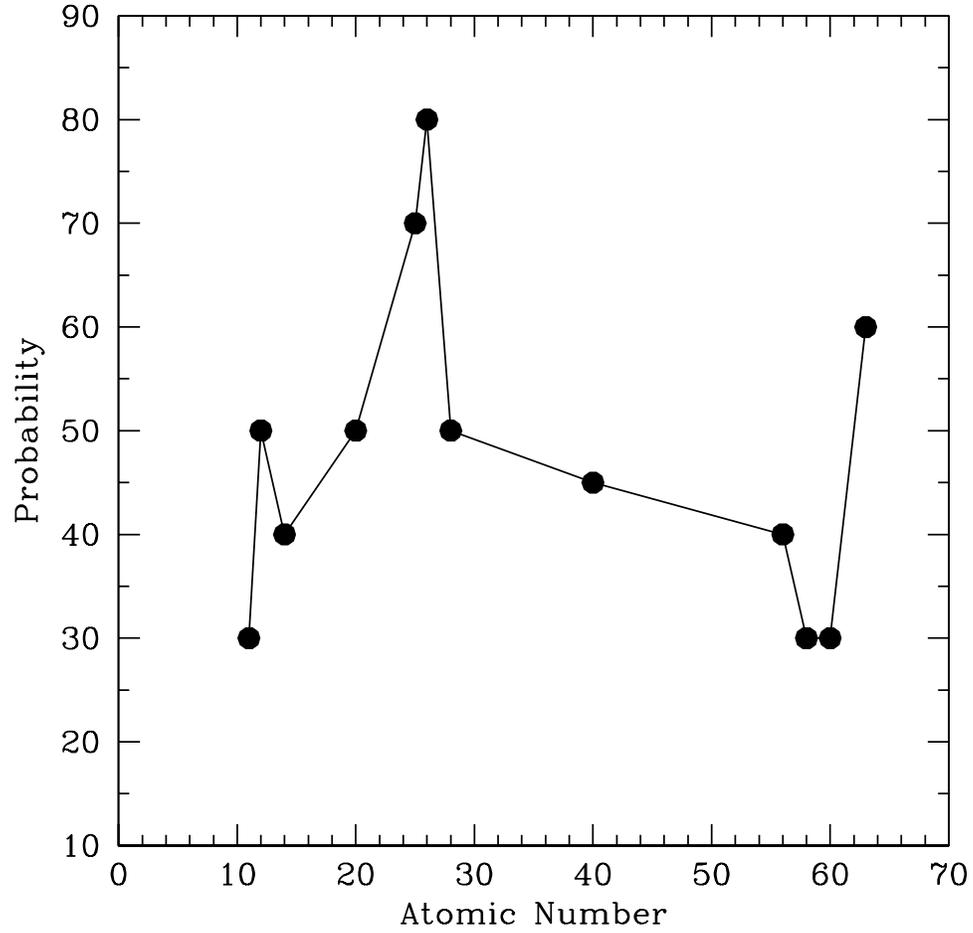}
\caption{The probability of finding a scatter as large as observed given the measuring errors and zero intrinsic scatter for the studied elements. Fe (N = 26) is the element most consistent with zero intrinsic scatter.}\label{probhr1614}
\end{center}
\end{figure}

Disregarding the four deviating stars, Table \ref{intscatter} summarizes the intrinsic star-to-star scatter $\sigma_{int}$ within the HR1614 moving group. By examining the uncertainties in our abundance error analysis we also calculate the uncertainty of the estimated intrinsic scatter. Assuming our abundance measurements follow a Gaussian distribution we estimate the confidence interval for the uncertainty in intrinsic scatter. Taking into account the sampling error in the observed scatter and a 10\% uncertainty in our adopted measurement errors, we find that the stated uncertainty limits in Table \ref{intscatter} are approximately 80\% confidence limits for $\sigma_{int}$. Further, based on a $\chi^{2}$ analysis we derive the probability of finding the observed scatter given the measurement errors and zero intrinsic scatter. Figure \ref{probhr1614} plots this probability for the different elements.\\


The above analysis does not include the four deviating stars. Since our aim is to obtain an estimate of the level of chemical homogeneity in 
all the moving group members, it is important for us to determine the nature of the
deviating stars. If they are found to be non-members or peculiar in some way (including binary systems, where the original birth abundance level may have been modified), then the level of homogeneity observed is indeed very high. If there is a reason to include the deviating stars, the mean level of homogeneity will rise to well over 0.1 dex. However given the clear bimodality of the results, it seems likely that these stars have a different origin. \\

\subsection{Stellar Membership}\label{sect:membership}

Following the bimodal nature of the chemical analysis results, one
immediately wonders if the deviating stars are non-members of the group. This
is much harder to establish for a moving group than for a bound cluster
system. Putting aside any chemical knowledge, space velocities and color-magnitude criteria are the key to establishing membership. If we are to use the tilted criteria of FH, we see in Figure \ref{fig:uv} that the deviating star HIP25840
lie outside their tilted boundary. Also adopting Eggen's criteria, this star is below the V-velocity clump. This velocity deviation would be sufficient to class this star as a non-member of the moving group. So we feel justified in excluding this star from further discussion. \\\\

Our radial velocity measurements found HIP13513 to have a variable radial
velocity. The red and blue spectra were observed a few weeks apart, and the
derived RV were greatly different. This prompted a search for
established binarity for these four deviating stars. Searching the Hipparcos
catalogue confirmed that stars HIP13513 and HIP6762 were binary stars. This would be a valid reason for us to discard these stars in our search for chemical homogeneity.\\

\begin{figure}
\begin{center}
\includegraphics[scale=0.7, angle=0]{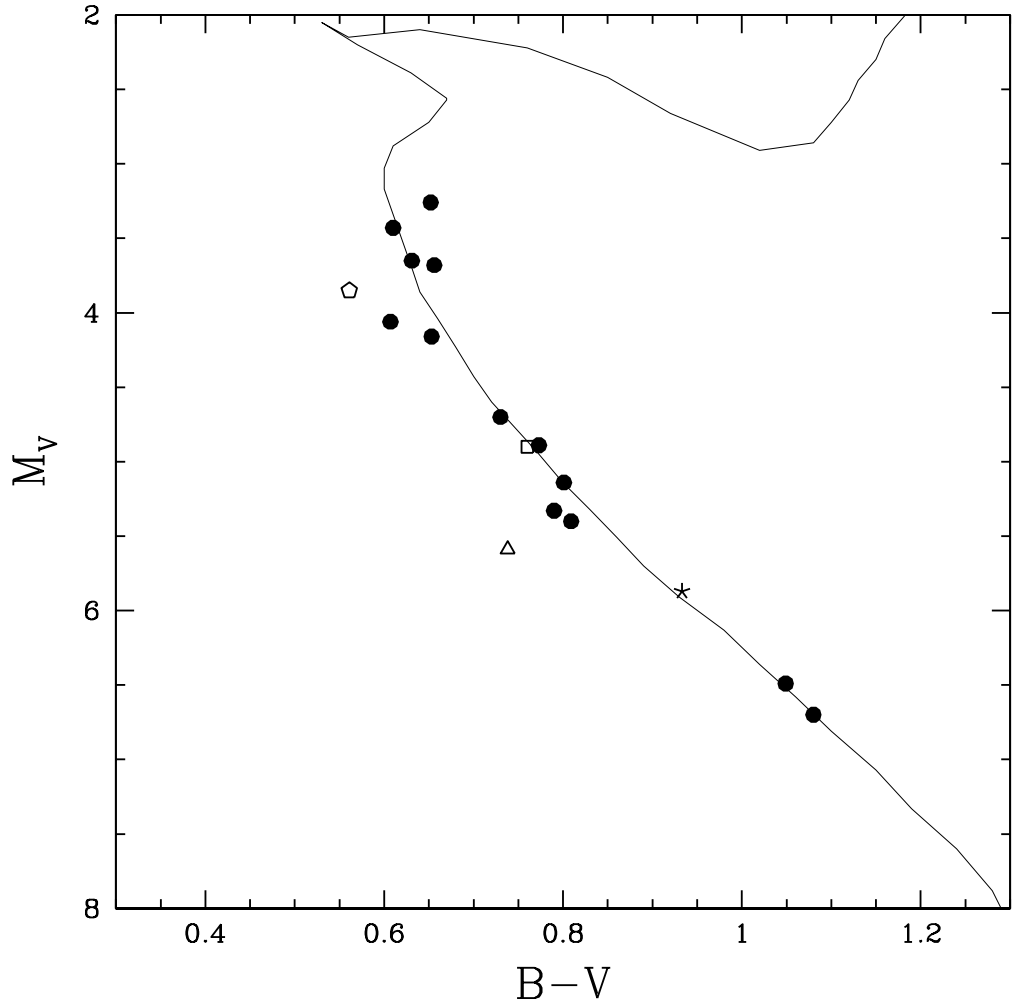}
\caption{V vs.~B-V color-magnitude diagram for our sample stars. The chemically
  deviating stars are identified with symbols as in Fig \ref{fig:fe}. The best fitting 2 Gyr \citet{bert} isochrone with Z = 0.05 has been overlaid to guide the eye.}\label{fig:cmd}
\end{center}
\end{figure}

The stars' position in the CMD also help identify possible non-members. Figure \ref{fig:cmd} shows the CMD of our sample stars with the
deviating stars marked with their respective symbols. HIP25840 and HIP6762 lie
clearly away from the main sequence, providing further reason to discard
these stars as possible non-members or binaries. HIP13513 happens to lie on
the main sequence, however given our earlier discussion on its variability, its
positioning on the CMD is likely to be by chance.\\

Our investigation fails to identify star HIP116970 as a non-member or
peculiar in any way. It satisfies both FH and Eggen's velocity criteria, lies in
the middle of the main sequence in the CMD, and has no variability or binary
observed in other surveys. 
A plausible scenario is that this star is a field star unassociated with the HR1614 moving group, but lies within the group's kinematical and photometric interval simply by chance. Out of our total sample of 18 stars we discover only the one star (if we disregard the other three stars assuming they are ``confirmed'' non-members or binaries) to be contaminating the HR1614 moving group. We are not aware of any literature calculations of the expected rate of contamination of moving groups. \\

In an attempt to estimate possible contamination, we obtained a synthetic stellar population based on Besancon models \citep{besancon} with stars limited to the magnitude, color, and velocity criteria of the HR1614 group stars. The stellar sample was restricted to stars lying along the main sequence in Figure \ref{fig:cmd} with 4 $<$ M$_{V} <$ 7 and 0.5 $<$ B-V $<$ 1.2, and width of $\Delta$(B-V) = $\pm$ 0.2. The velocity space was restricted to stars lying along the tilt in Figure \ref{fig:uv} with -55 $<$ U $<$ 25 \kms\ and -70 $<$ V $<$ -40 \kms, and width $\Delta$V = $\pm$ 20 \kms. In this selected sample 15\% of stars had solar-level metallicities as defined by [Fe/H] = 0.00 $\pm$ 0.05. No prior restrictions were made on the metallicity. This suggests that one in seven stars from the background galactic disk in this restricted interval of absolute magnitude, color and velocity will have solar-level metallicity [Fe/H] = 0.00 $\pm$ 0.05.  We can therefore expect about one in seven such background stars in our HR 1614 moving group sample.


\subsection{Implications for Chemical Tagging}

The clear bimodal abundance pattern seen for the $\alpha$, Fe, Fe-peak and light s-process elements is interesting. As discussed above, of the four
deviating stars, only one cannot be classed as a non-member, a binary or otherwise peculiar star based on its kinematics or photometric properties. Although we are not aware of any prior calculation of the expected rate of contamination from field stars that are not part of the group, our simple estimate based on Besancon stellar population models show that some contamination (15\%) is to be expected. It is likely that there are other such contaminating stars in this (and other) moving groups, that are not yet identified due to the lack of detailed chemical information. Our results are a clear demonstration of the necessity and importance of obtaining detailed chemical information, and that chemical probing is essential to see the true story behind the history of any stellar stream in the Galactic disk.\\

Our results support the conclusion of \citet{FH} regarding the reality of the 
HR 1614 moving group. Several authors have argued that at least some 
moving stellar groups are not the dispersed remnants of star-forming events, 
but rather are dynamical in origin, resulting from the dynamical effects
of the galactic bar or spiral structure \citep[eg.][]{dehnen99,desimone,famaey}. We would not expect
such dynamical groups to be coeval and chemically homogeneous.  The 
homogeneity of the HR1614 group in age and abundance supports the view 
that this group is indeed the remnant of a dispersed star-forming event. 
Its kinematical coherence shows that such a dispersing system need not be 
significantly perturbed by external dynamical influences like galactic spiral 
structure or giant molecular clouds, at least over a period of order 2 Gyr.\\

Finally, the high level of chemical homogeneity observed across the confirmed HR1614 moving group members is a major step forward for chemical tagging. Earlier \citet{desilva} reported chemical homogeneity in the Hyades which is a younger bound star cluster. Our study on the very old open cluster Cr 261 (De Silva et al., in preparation) also show similar levels of homogeneity and in it we further explore the implications of our combined results on chemical tagging. Here we have shown that an unbound intermediate-aged cluster is also chemically identifiable. Although much work is yet to be done, the chemical tagging technique and the prospect of unravelling the dissipative history of the Galactic disk now seems viable.\\



\acknowledgments
GMD would like to thank Mary Williams for her assistance with determining radial and space velocities, and Agris Kalnajs for useful discussions on the dynamics of stellar streams. We thank the anonymous referee for his helpful comments. This research has made use of the Vienna Atomic Line Database, operated at Vienna, Austria, and the Database on Rare Earths At Mons University, operated at Mons, Belgium.

\clearpage

\begin{deluxetable}{lcccccccc}
\tabletypesize{\normalsize}
\tablecaption{Stellar Sample\label{tab:sample}
}
\tablewidth{0pt}
\tablehead{
\colhead{HIP}& 
\colhead{HD} & \colhead{RA} & \colhead{DEC} & \colhead{V} & 
\colhead{B-V} & \colhead{V-K\tablenotemark{a}} & \colhead{Source\tablenotemark{b}}

}

\startdata

23311  & 32147  & 05 00 48.68 & -05 45 03.5 & 6.22 & 1.05 & 2.51 & E,FH\\ 
110996 & 213042 & 22 29 15.23 & -30 01 06.3 & 7.65 & 1.08 & 2.53 & E \\
26834  & 37986  & 05 41 53.54 & -15 37 48.9 & 7.36 & 0.80 & 1.73 & FH \\
109378 & 210277 & 22 09 29.82 & -07 32 51.2 & 6.54 & 0.77 & 1.74 & E,FH \\
22940  & 31452  & 04 56 10.61 & -05 40 24.4 & 8.43 & 0.84 & 1.96 & FH \\
116554 & 222013 & 23 37 15.31 & -45 28 30.8 & 9.22 & 0.81 & 1.88 & E,FH \\
10599  & 13997  & 02 16 27.60 & +12 22 49.1 & 7.99 & 0.79 & 1.76 & E,FH \\
110843 & 212708 & 22 27 24.38 & -49 21 54.5 & 7.48 & 0.73 & 1.68 & FH \\
17960  & 24040  & 03 50 22.90 & +17 28 37.1 & 7.50 & 0.65 & 1.53 & FH \\
22336  & 30562  & 04 48 36.20 & -05 40 24.4 & 5.77 & 0.63 & 1.46 & FH \\
102393 & 197623 & 20 44 57.03 & +00 17 31.7 & 7.55 & 0.66 & 1.52 & FH \\
11575  & 15590  & 02 29 11.90 & -42 04 31.1 & 7.98 & 0.65 & 1.47 & E \\
9353   & 12235  & 02 00 09.16 & +03 05 49.2 & 5.89 & 0.61 & 1.40 & E,FH \\
102018 & 196800 & 20 40 22.33 & -24 07 04.9 & 7.21 & 0.61 & 1.42 & FH \\
13513  & 18168  & 02 54 02.78 & -35 54 16.8 & 8.23 & 0.93 & 2.36 & FH \\
6762   & 8828   & 01 27 01.55 & -00 09 27.3 & 7.96 & 0.74 & 1.80 & E \\
116970 & 222655 & 23 42 42.22 & -41 14 51.3 & 9.57 & 0.76 & 1.74 & E \\
25840  & 36379  & 05 30 59.85 & -10 04 49.1 & 6.94 & 0.56 & 1.39 & E\\

\enddata
\tablenotetext{a}{K magnitudes from 2MASS}
\tablenotetext{b}{FH = \citet{FH}, E = \citet{e98}}
\end{deluxetable}

\begin{deluxetable}{lccrcc|lccrcc|lccrcc} 
\tabletypesize{\scriptsize}
\rotate
\tablecolumns{12} 
\tablewidth{0pt} 
\tablecaption{Line list\label{hr1614:lines}}
\tablehead{ 
\colhead{Wavelength(\AA)} &
\colhead{Species} &
\colhead{LEP(eV)} &
\colhead{log $gf$} &
\colhead{EW(m\AA)} &
\colhead{} &
\colhead{Wavelength(\AA)} &
\colhead{Species} &
\colhead{LEP(eV)} &
\colhead{log $gf$} &
\colhead{EW(m\AA)} &
\colhead{} &
\colhead{Wavelength(\AA)} &
\colhead{Species} &
\colhead{LEP(eV)} &
\colhead{log $gf$}&
\colhead{EW(m\AA)}  
}
\startdata
5688.19 & Na {\sc i} & 2.11 & $-$0.420 & 265.0 & & 5618.63 & Fe {\sc i} & 4.21 & $-$1.292 & 75.0 & & 4582.83 & Fe {\sc ii}& 2.84 & $-$3.094 & 61.2 \\
6154.23 & Na {\sc i} & 2.10 & $-$1.530 & 128.0 & & 5705.47 & Fe {\sc i} & 4.30 & $-$1.420 & 59.6 & & 4620.52 & Fe {\sc ii}& 2.83 & $-$3.079 & 39.0\\
6160.75 & Na {\sc i} & 2.10 & $-$1.230 & 140.0 & & 5741.85 & Fe {\sc i} & 4.25 & $-$1.689 & 64.0 & & 4670.17 & Fe {\sc ii}& 2.58 & $-$3.904 & 26.5\\
5711.09 & Mg {\sc i} & 4.35 & $-$1.833 & 171.0 & & 5778.45 & Fe {\sc i} & 2.59 & $-$3.480 & 58.9 & & 5991.38 & Fe {\sc ii}& 3.15 & $-$3.557 & 19.2\\
6318.72 & Mg {\sc i} & 5.11 & $-$1.970 &  79.0 & & 5811.92 & Fe {\sc i} & 4.14 & $-$2.430 & 25.5 & & 6084.11 & Fe {\sc ii}& 3.20 & $-$3.808 & 10.1\\
6696.02 & Al {\sc i} & 3.14 & $-$1.340 & 110.0 & & 5837.70 & Fe {\sc i} & 4.29 & $-$2.340 & 29.0 & & 6149.26 & Fe {\sc ii}& 3.89 & $-$2.724 & 16.7\\
6698.67 & Al {\sc i} & 3.14 & $-$1.640 &  84.3 & & 5853.16 & Fe {\sc i} & 1.49 & $-$5.280 & 41.6 & & 6247.56 & Fe {\sc ii}& 3.89 & $-$2.329 & 22.3\\
7835.31 & Al {\sc i} & 4.02 & $-$0.470 & 125.0 & & 5856.10 & Fe {\sc i} & 4.29 & $-$1.640 & 60.5 & & 6416.92 & Fe {\sc ii}& 3.89 & $-$2.740 & 37.0\\
7836.13 & Al {\sc i} & 4.02 & $-$0.310 & 148.8 & & 5858.79 & Fe {\sc i} & 4.22 & $-$2.260 & 28.1 & & 6432.68 & Fe {\sc ii}& 2.89 & $-$3.708 & 28.9\\ 
5665.56 & Si {\sc i} & 4.92 & $-$1.940 &  50.5 & & 5927.80 & Fe {\sc i} & 4.65 & $-$1.090 & 64.5 & & 6456.38 & Fe {\sc ii}& 3.90 & $-$2.075 & 30.8\\ 
5684.49 & Si {\sc i} & 4.95 & $-$1.550 &  67.0 & & 5956.69 & Fe {\sc i} & 0.86 & $-$4.608 & 102.5& & 7224.49 & Fe {\sc ii}& 3.89 & $-$3.243 & 17.5\\
5690.43 & Si {\sc i} & 4.93 & $-$1.770 &  55.0 & & 6054.08 & Fe {\sc i} & 4.37 & $-$2.310 & 26.0 & & 7711.72 & Fe {\sc ii}& 3.90 & $-$2.543 & 33.5\\
5948.54 & Si {\sc i} & 5.08 & $-$1.230 &  90.6 & & 6120.24 & Fe {\sc i} & 0.91 & $-$5.970 & 40.0 & & 5846.99 & Ni {\sc i} & 1.68 & $-$3.210 & 70.5\\
6142.48 & Si {\sc i} & 5.62 & $-$1.540 &  32.8 & & 6151.62 & Fe {\sc i} & 2.17 & $-$3.299 & 83.0 & & 6086.28 & Ni {\sc i} & 4.26 & $-$0.515 & 65.7\\
6145.01 & Si {\sc i} & 5.62 & $-$1.362 &  40.5 & & 6157.73 & Fe {\sc i} & 4.08 & $-$1.320 & 87.4 & & 6175.37 & Ni {\sc i} & 4.09 & $-$0.535 & 68.2\\
6155.13 & Si {\sc i} & 5.62 & $-$0.786 &  83.2 & & 6180.20 & Fe {\sc i} & 2.73 & $-$2.637 & 101.5& & 6177.24 & Ni {\sc i} & 1.83 & $-$3.510 & 46.8\\
5868.57 & Ca {\sc i} & 2.93 & $-$1.570 &  79.5 & & 6200.31 & Fe {\sc i} & 2.61 & $-$2.437 & 120.0& & 6204.60 & Ni {\sc i} & 4.09 & $-$1.140 & 42.5\\
6169.04 & Ca {\sc i} & 2.52 & $-$0.797 & 211.5 & & 6219.28 & Fe {\sc i} & 2.20 & $-$2.433 & 148.5& & 6635.12 & Ni {\sc i} & 4.42 & $-$0.828 & 40.0\\
6169.56 & Ca {\sc i} & 2.53 & $-$0.478 & 261.0 & & 6229.23 & Fe {\sc i} & 2.84 & $-$2.846 & 69.0 & & 6772.32 & Ni {\sc i} & 3.66 & $-$0.987 & 73.8\\
6455.60 & Ca {\sc i} & 2.52 & $-$1.290 & 121.2 & & 6246.32 & Fe {\sc i} & 3.60 & $-$0.894 & 215.0& & 4050.33 & Zr {\sc ii}& 0.71 & $-$1.000 & 33.0\\
6572.80 & Ca {\sc i} & 0.00 & $-$4.280 & 115.0 & & 6265.13 & Fe {\sc i} & 2.17 & $-$2.550 & 149.0& & 5853.69 & Ba {\sc ii}& 0.60 & $-$1.000 & 75.0\\
6013.53 & Mn {\sc i} & 3.07 & $-$0.251 & 168.7 & & 6270.22 & Fe {\sc i} & 2.86 & $-$2.500 & 84.5 & & 6141.73 & Ba {\sc ii}& 0.70 & $-$0.076 & 148.0 \\
6016.67 & Mn {\sc i} & 3.08 & $-$0.100 & 170.6 & & 6271.28 & Fe {\sc i} & 3.33 & $-$2.728 & 57.8 & & 6496.91 & Ba {\sc ii}& 0.60 & $-$0.410 & 115.5\\
6021.80 & Mn {\sc i} & 3.08 &    0.034 & 176.8 & & 6297.79 & Fe {\sc i} & 2.22 & $-$2.740 & 114.5& & 4083.22 & Ce {\sc ii}& 0.70 &    0.270 & \nodata\\
4139.93 & Fe {\sc i} & 0.99 & $-$3.629 & 126.0 & & 6336.82 & Fe {\sc i} & 3.68 & $-$0.916 & 193.0& & 4137.65 & Ce {\sc ii}& 0.52 &    0.440 & \nodata\\
4216.19 & Fe {\sc i} & 0.00 & $-$3.356 & 400.0 & & 6411.65 & Fe {\sc i} & 3.65 & $-$0.734 & 232.5& & 4364.65 & Ce {\sc ii}& 0.49 &    0.230 & \nodata\\
4232.72 & Fe {\sc i} & 0.11 & $-$4.928 & 121.0 & & 6574.23 & Fe {\sc i} & 0.99 & $-$5.004 & 83.0 & & 4562.36 & Ce {\sc ii}& 0.48 &    0.230 & \nodata\\
4347.24 & Fe {\sc i} & 0.00 & $-$5.503 &  83.5 & & 6609.11 & Fe {\sc i} & 2.56 & $-$2.692 & 118.0& & 4628.16 & Ce {\sc ii}& 0.52 &    0.200 & \nodata\\
4389.24 & Fe {\sc i} & 0.05 & $-$4.583 & 116.0 & & 6699.16 & Fe {\sc i} & 4.59 & $-$2.170 & 20.5 & & 4012.24 & Nd {\sc ii}& 0.63 &    0.810 & \nodata\\
4439.88 & Fe {\sc i} & 2.28 & $-$3.002 &  70.5 & & 6739.52 & Fe {\sc i} & 1.56 & $-$4.820 & 49.0 & & 4012.69 & Nd {\sc ii}& 0.00 & $-$0.600 & \nodata\\
4442.34 & Fe {\sc i} & 2.19 & $-$1.255 & 339.5 & & 6810.26 & Fe {\sc i} & 4.60 & $-$1.000 & 80.3 & & 4068.89 & Nd {\sc ii}& 0.00 & $-$1.420 & \nodata\\ 
4445.48 & Fe {\sc i} & 0.08 & $-$5.441 &  85.0 & & 4178.86 & Fe {\sc ii}& 2.57 & $-$2.720 & 87.0 & & 4462.98 & Nd {\sc ii}& 0.56 &    0.040 & \nodata\\
4489.74 & Fe {\sc i} & 0.12 & $-$3.966 & 189.0 & & 4491.41 & Fe {\sc ii}& 2.86 & $-$2.684 & 63.0 & & 4129.23 & Eu {\sc ii}& 0.00 &    0.270 & \nodata\\
4531.15 & Fe {\sc i} & 1.48 & $-$2.155 & 295.0 & & 4508.28 & Fe {\sc ii}& 2.86 & $-$2.312 & 64.5 & & \nodata & \nodata& \nodata& \nodata & \nodata \\
4602.01 & Fe {\sc i} & 1.61 & $-$3.154 & 108.6 & & 4576.33 & Fe {\sc ii}& 2.84 & $-$2.822 & 48.0 & & \nodata & \nodata& \nodata& \nodata & \nodata

\enddata

\end{deluxetable}

\begin{deluxetable}{lccccc}
\tabletypesize{\normalsize}
\tablecaption{Adopted Stellar Parameters\label{tab:params}
}
\tablewidth{0pt}
\tablehead{
\colhead{HIP} & 
\colhead{Mass (M$_{\odot}$)} & \colhead{M$_{\mbox{bol}}$} & \colhead{T$_{eff}$(K)}
& \colhead{log $g$(cm~s$^{-2}$)} & \colhead{$\xi$(\kms)} 
}

\startdata

23311  & 0.80 &6.34 & 4850 & 4.4 & 0.8 \\ 
110996 & 0.78 &6.34 & 4800 & 4.3 & 0.8 \\
26834  & 0.95 &5.33 & 5500 & 4.3 & 0.7 \\
109378 & 0.97 &5.25 & 5500 & 4.3 & 0.7 \\
22940  & 0.93 &5.50 & 5250 & 4.5 & 0.4 \\
116554 & 0.92 &5.50 & 5400 & 4.7 & 0.6 \\
10599  & 0.95 &5.33 & 5450 & 4.5 & 1.0 \\
110843 & 1.00 &5.00 & 5600 & 4.3 & 1.1 \\
17960  & 1.12 &4.32 & 5800 & 4.4 & 1.1 \\
22336  & 1.25 &3.70 & 5900 & 4.4 & 1.3 \\
102393 & 1.20 &3.80 & 5900 & 4.4 & 1.1 \\
11575  & 1.30 &3.20 & 5900 & 4.1 & 1.2 \\
9353   & 1.30 &3.45 & 6000 & 4.2 & 1.2 \\
102018 & 1.20 &4.14 & 6000 & 4.4 & 1.2 \\
13513  & 0.87 &5.90 & 5050 & 4.5 & 1.1 \\
6762   & 0.95 &5.17 & 5400 & 4.6 & 0.9 \\
116970 & 0.97 &5.17 & 5500 & 4.7 & 0.9 \\
25840  & 1.25 &3.85 & 6150 & 4.5 & 1.2 \\

\enddata

\end{deluxetable}

\clearpage

\begin{deluxetable}{lccccccccccccccccc}
\tabletypesize{\normalsize}
\rotate
\tablecaption{Absolute Abundances (log $\epsilon$)\label{tab:abundance}
}
\tablewidth{0pt}
\tablehead{
\colhead{HIP ID}& 
\colhead{Na} & \colhead{Mg} & \colhead{Al} & \colhead{Si} & 
\colhead{Ca} & \colhead{Mn} & \colhead{Fe}
& \colhead{Ni} & \colhead{Zr} & \colhead{Ba} & \colhead{Ce}
& \colhead{Nd} & \colhead{Eu}
}

\startdata

23311  & 6.56 &7.86 & 6.77 & 7.78 & 6.65 & 5.79 & 7.77 & 6.63 & 2.72 & 2.52 & 1.75 & 1.58 & 0.74\\ 
110996 & 6.61 &7.83 & 6.73 & 7.77 & 6.67 & 5.79 & 7.75 & 6.57 & 2.65 & 2.51 & 1.68 & 1.53 & 0.71 \\
26834  & 6.62 &7.98 & 6.72 & 7.90 & 6.71 & 5.82 & 7.83 & 6.69 & 2.77 & 2.67 & 1.74 & 1.56 & 0.71 \\
109378 & 6.50 &7.92 & 6.65 & 7.83 & 6.60 & 5.72 & 7.76 & 6.55 & 2.77 & 2.61 & 1.73 & 1.56 & 0.76 \\
22940  & 6.41 &7.87 & 6.60 & 7.79 & 6.60 & 5.76 & 7.75 & 6.63 & 2.94 & 2.59 & 1.73 & 1.58 & 0.73 \\
116554 & 6.35 &7.78 & 6.52 & 7.77 & 6.55 & 5.73 & 7.75 & 6.54 & 2.73 & 2.56 & 1.73 & 1.62 & 0.71 \\
10599  & 6.54 &7.82 & 6.68 & 7.88 & 6.59 & 5.73 & 7.77 & 6.64 & 2.83 & 2.52 & 1.72 & 1.57 & 0.68 \\
110843 & 6.55 &7.84 & 6.65 & 7.83 & 6.59 & 5.76 & 7.79 & 6.58 & 2.73 & 2.51 & 1.71 & 1.55 & 0.71\\
179690 & 6.49 &7.87 & 6.64 & 7.78 & 6.58 & 5.72 & 7.72 & 6.59 & 2.72 & 2.55 & 1.75 & 1.56 & 0.74 \\
22336  & 6.48 &7.78 & 6.59 & 7.83 & 6.56 & 5.74 & 7.76 & 6.57 & 2.77 & 2.62 & 1.73 & 1.56 & 0.73 \\
102393 & 6.57 &7.80 & 6.64 & 7.88 & 6.61 & 5.76 & 7.82 & 6.63 & 2.79 & 2.60 & 1.74 & 1.53 & 0.76 \\
11575  & 6.58 &7.80 & 6.57 & 7.81 & 6.55 & 5.73 & 7.77 & 6.56 & 2.75 & 2.60 & 1.68 & 1.56 & 0.76 \\
9353   & 6.61 &7.76 & 6.61 & 7.87 & 6.57 & 5.74 & 7.73 & 6.57 & 2.72 & 2.53 & 1.67 & 1.56 & 0.74 \\
102018 & 6.51 &7.76 & 6.51 & 7.76 & 6.59 & 5.73 & 7.72 & 6.53 & 2.73 & 2.51 & 1.64 & 1.55 & 0.78 \\
13513  & 6.17 &7.61 & 6.34 & 7.59 & 6.38 & 5.39 & 7.55 & 6.31 & 2.44 & 2.33 & 1.55 & 1.43 & 0.61 \\
6762   & 6.06 &7.47 & 6.24 & 7.53 & 6.25 & 5.32 & 7.47 & 6.21 & 2.51 & 2.41 & 1.61 & 1.41 & 0.61 \\
116970 & 6.20 &7.63 & 6.27 & 7.61 & 6.38 & 5.38 & 7.55 & 6.32 & 2.50 & 2.46 & 1.67 & 1.44 & 0.65 \\
25840  & 6.16 &7.61 & 6.24 & 7.47 & 6.27 & 5.36 & 7.44 & 6.21 & 2.51 & 2.42 & 1.59 & 1.42 & 0.65\\

\enddata

\end{deluxetable}





\begin{deluxetable}{lccccccccccccccccccccccccc}
\tabletypesize{\scriptsize}
\rotate
\tablecaption{Abundance Sensitivities \label{tab:sens}
}
\tablewidth{0pt}
\tablehead{
\colhead{Example Star}& \colhead{Model Parameter}&
\colhead{Na} & \colhead{Mg} & \colhead{Al} & \colhead{Si} &
\colhead{Ca} & \colhead{Mn} & \colhead{Fe}
& \colhead{Ni} & \colhead{Zr} & \colhead{Ba} & \colhead{Ce}
& \colhead{Nd} & \colhead{Eu}
}

\startdata

HR1614 &T$_{\rm eff}\pm$50 & $\pm$0.04 & $\pm$0.01 & $\pm$0.03 & $\mp$0.02 & $\pm$0.03 & 0.00 & $\pm$0.02 & $\mp$0.01 & $\pm$0.01 & $\pm$0.01 & $\pm$0.02 & $\pm$0.01 & $\pm$0.01 \\
($T_{\rm eff}$ = 4850K)&log $g \pm$0.1&$\mp$0.05 & $\mp$0.02 & $\mp$0.04 & $\pm$0.01 & $\mp$0.04 & 0.00 & 0.00 & $\pm$0.01 & $\pm$0.04 & 0.00 & 0.00 & 0.00 & $\pm$0.03 \\
&$\xi\pm$0.1 &$\mp$0.01 & $\mp$0.02 & $\mp$0.02 & $\mp$0.01 & $\mp$0.04 & $\pm$0.02 & $\mp$0.04 & $\mp$0.03 & $\mp$0.05 & $\mp$0.05 & $\pm$0.01 & $\pm$0.02 & 0.00  \\
&$\Delta$ EW / synth &$\pm$0.03 & $\pm$0.06 & $\pm$0.05 & $\pm$0.04 & $\pm$0.03 & $\pm$0.03 & $\pm$0.03 & $\pm$0.03 & $\pm$0.02 & $\pm$0.02 & $\pm$0.02 & $\pm$0.02 & $\pm$0.02  \\

\hline

HIP9353 &T$_{\rm eff}\pm$50 & $\pm$0.02 & $\pm$0.02 & $\pm$0.02 & $\mp$0.01 & $\pm$0.02 & $\pm$0.01 & $\pm$0.04 & $\pm$0.01 & 0.00 & $\pm$0.01 & $\pm$0.01 &0.00 & $\pm$0.01 \\
($T_{\rm eff}$ = 6000K)&log $g \pm$0.1&$\mp$0.02 & $\mp$0.01 & $\mp$0.01 & 0.00 & $\mp$0.01 & 0.00 & $\mp$0.01 & 0.00 & $\pm$0.04 & $\pm$0.01 & $\pm$0.01 & $\pm$0.02 & $\pm$0.03 \\
&$\xi\pm$0.1 &$\mp$0.02 & $\mp$0.02 & $\mp$0.01 & $\mp$0.03 & $\mp$0.04 & $\pm$0.02 & $\mp$0.03 & $\mp$0.04 & $\mp$0.04 & $\mp$0.05 & 0.00 & 0.00 & $\pm$0.01 \\
&$\Delta$ EW /synth&$\pm$0.04 & $\pm$0.05 & $\pm$0.06 & $\pm$0.04 & $\pm$0.02 & $\pm$0.03 & $\pm$0.01 & $\pm$0.04 & $\pm$0.02 & $\pm$0.02 & $\pm$0.02 & $\pm$0.02 & $\pm$0.02 \\

\enddata
\end{deluxetable}

\begin{deluxetable}{ccccccccc}
\tabletypesize{\normalsize}
\tablecaption{Radial Velocities\label{tab:rv}
}
\tablewidth{0pt}
\tablehead{
\colhead{HIP ID} & \colhead{This study} & \colhead{N04} &
\colhead{FH sources\tablenotemark{*} } 
}

\startdata
 23311  &  21.6 &  21.0  &   27.0 \\
110996  &   5.3 &   4.8  &    7.4 \\
 26834  &  59.1 &\nodata &   67.0 \\
109378  & -21.3 &\nodata &  -24.1 \\
 22940  &  15.2 &\nodata &   14.3 \\
116554  & -16.5 & -15.1  &   -2.6 \\
 10599  & -21.5 & -21.3  &  -23.6 \\
110843  &   4.7 &   4.9  &   -7.3 \\
 17960  & -10.1 &\nodata &   -7.6 \\
 22336  &  76.7 &  77.0  &   78.6 \\
102393  & -69.2 &\nodata &  -71.0 \\
 11575  &  37.0 &  36.9  &   36.8 \\
  9353  & -18.1 & -18.5  &  -18.4 \\
102018  & -63.5 & -63.8  &  -61.2 \\ 
 13513  &  3.5/-46.9 &   4.0  &    3.0 \\
  6762  &  13.7 &  13.2  &    3.5 \\
116970  &  27.7 &\nodata &   26.9 \\
 25840  &  22.9 &  22.8  &   \nodata  \\

\hline
\vspace{0.2cm}\\

\hline
RV standards: && This study  &  \citet{maurice}\\
&HD6677  &  19.7 &  19.5 \\
&CPD432527& 20.0 &  19.1\\

\enddata
\tablenotetext{*}{\citet{HIC,barbier,grenier}}
\end{deluxetable}

\begin{deluxetable}{lccccccccccc}
\tabletypesize{\normalsize}
\tablecaption{UVW Velocities\label{tab:uvw}
}
\tablewidth{0pt}
\tablehead{
\colhead{HIP ID} & \colhead{U} &
\colhead{V} & \colhead{W} &
\colhead{U(FH)} & \colhead{V(FH)} & \colhead{W(FH)}

}

\startdata
 23311 &  14.9  &  -49.7 &  -3.8  &   10.65 & -51.83 &  -6.33 \\
110996 &  15.6  &  -54.7 &  -7.8 &   16.55 & -54.51 &  -9.70 \\
 26834 & -21.6  &  -56.3 &  6.2   &  -27.11 & -61.11 &   3.02 \\
109378 &  13.9  &  -45.4 &   2.9  & 12.46 & -46.78 &   3.09 \\
 22940 &  15.2  &  -56.7 &  -3.1  &  15.40 & -56.50 &  -3.05 \\
116554 &   1.0  &  -53.3 &   28.1 &    3.27 & -54.08 &  17.25 \\
 10599 &  10.9  &  -50.2 &   11.9 &   12.00 & -50.73 &  13.64 \\
110843 &  -29.3 &  -62.1 &  -12.7 &  -35.96 & -60.25 &  -2.93 \\
 17960 &  21.1  &  -54.2 &   -7.3 &   20.26 & -53.93 &  -7.57 \\
 22336 & -41.8  &  -67.7 &  -13.8 &  -43.26 & -68.33 & -14.98 \\
102393 & -45.3  &  -54.4 &  -3.4  & -47.02 &  -55.19 &  -3.02 \\
 11575 &  14.7  &  -51.7 &    6.6 &  \nodata&\nodata &\nodata \\
  9353 &  11.7  &  -48.6 &   12.3 &   11.55 & -48.59 &  12.43 \\
102018 & -50.6  &  -66.9 &  -1.4  &  -48.95 & -65.96 &  -2.84 \\
 13513 &-14.5/-3.1& -59.7/-39.8 & 38.6/83.3 &  -14.75 & -59.61 &  39.00 \\ 
  6762 & -3.9 &  -53.8 &  -22.1 &   -4.09 & -53.95 & -22.13 \\
116970 &   9.9  &  -55.7 &  -16.2 & 9.87 & -55.64 & -15.42 \\ 
 25840 &  19.8  &  -70.6 &  -7.0  &  \nodata&\nodata &\nodata \\

\enddata

\end{deluxetable}

\begin{deluxetable}{cccccc}
\tabletypesize{\normalsize}
\tablecaption{Mean Abundances and Scatter\label{tab:scatter}
}
\tablewidth{0pt}
\tablehead{
\colhead{Element} & \colhead{$\langle$Members$\rangle$} & \colhead{$\sigma$} & \colhead{$\langle$All Stars$\rangle$} & \colhead{$\sigma$}  
}

\startdata

Fe & 7.77 & 0.033 & 7.71 & 0.117 \\
Na & 6.53 & 0.078 & 6.44 & 0.178 \\
Mg & 7.83 & 0.063 & 7.77 & 0.125 \\
Al & 6.63 & 0.075 & 6.55 & 0.169 \\
Si & 7.82 & 0.047 & 7.76 & 0.126 \\
Ca & 6.60 & 0.046 & 6.53 & 0.130 \\
Mn & 5.75 & 0.030 & 5.67 & 0.169 \\
Ni & 6.59 & 0.046 & 6.52 & 0.148 \\
Zr & 2.76 & 0.067 & 2.70 & 0.129 \\
Ba & 2.56 & 0.051 & 2.52 & 0.085 \\
Ce & 1.71 & 0.034 & 1.69 & 0.059 \\
Nd & 1.56 & 0.022 & 1.53 & 0.062 \\
Eu & 0.73 & 0.027 & 0.71 & 0.051 \\

\enddata

\end{deluxetable}

\begin{deluxetable}{cccccc}
\tabletypesize{\normalsize}
\tablecaption{Intrinsic scatter\label{intscatter}
}
\tablewidth{0pt}
\tablehead{
\colhead{Element} & \colhead{$\sigma_{int}$} & \colhead{uncertainty}  
}

\startdata

Fe & 0.000  & 0.010 \\
Na & 0.034  & 0.050 \\
Mg & 0.000  & 0.020 \\
Al & 0.017  & 0.040 \\
Si & 0.000  & 0.017 \\
Ca & 0.000  & 0.020 \\
Mn & 0.000  & 0.010 \\
Ni & 0.009  & 0.010 \\
Zr & 0.000  & 0.030 \\
Ba & 0.000  & 0.025 \\
Ce & 0.016  & 0.020 \\
Nd & 0.000  & 0.010 \\
Eu & 0.000  & 0.018 \\

\enddata

\end{deluxetable}

\clearpage

\end{document}